\documentclass[aps,prl,preprint,superscriptaddress]{revtex4-1}

\usepackage{amsmath}
\usepackage{amssymb}
\usepackage{amstext}
\usepackage{bm}
\usepackage{graphicx}
\usepackage{multirow}
\usepackage{color}
\usepackage{mathtools}
\usepackage{float}

\begin{document}
\title[]{Probing the limits of vortex mode generation and detection with spatial light modulators}

\author{Jonathan Pinnell}
\affiliation{School of Physics, University of the Witwatersrand, Johannesburg 2000, South Africa}
\author{Valeria Rodr\'{i}guez-Fajardo}
\affiliation{School of Physics, University of the Witwatersrand, Johannesburg 2000, South Africa}
\author{Andrew Forbes}
\affiliation{School of Physics, University of the Witwatersrand, Johannesburg 2000, South Africa}
\affiliation{Corresponding author: Andrew.Forbes@wits.ac.za}

\begin{abstract}
\noindent Spatial light modulators (SLMs) are popular tools for generating structured light fields and have fostered numerous applications in optics and photonics. Here, we explore the limits of what fields these devices are capable of generating and detecting in the context of so-called vortex beams carrying orbital angular momentum (OAM). Our main contributions are to quantify (theoretically and experimentally) how the pixelation of the SLM screen affects the quality of the generated vortex mode and to offer useful heuristics on how to optimise the performance of the displayed digital hologram. In so doing, we successfully generate and detect a very high order optical vortex mode with topological charge $\ell = 600$, the highest achieved to date using SLMs. Since the OAM degree of freedom is frequently touted as offering a potentially unbounded state space, we hope that this work will inspire researchers to make more use of higher order vortex modes.
\end{abstract}
\maketitle

\section{Introduction}

\noindent Spatial light modulators (SLMs) employing either liquid crystal displays or digital micro-mirror arrays have recently become ubiquitous tools in the modern optical laboratory owing to the versatility and ease-of-use that they provide for the on-demand creation, detection and manipulation of structured light fields \cite{Forbes2016}. In contrast to traditional holograms, which are static and time consuming to develop, SLMs have provided experimenters with the ability to generate reconfigurable digital holograms which are displayed on a relatively small but high resolution screen. Owing to this and the fact that very little specialised knowledge is required to utilise such devices, their popularity amongst researchers and industry professionals in the field of optics and photonics has increased dramatically \cite{lazarev2019beyond}. Applications of SLMs abound, ranging from communications \cite{wang2017data} and microscopy \cite{maurer2011spatial} to quantum information processing \cite{Mirhosseini2015}, metrology \cite{Lazarev2012} and optical manipulation \cite{grier2003revolution}. The reconfigurable nature of SLMs at moderately high refresh rates have meant that they are favoured in applications where static approaches were previously used. Further, they have facilitated the development of digital optical characterisation in real-time, such as modal decomposition \cite{Flamm2012B} and the determination of beam quality factor \cite{Schmidt2011real}, among others. 

As their name suggests, SLMs are able to modulate the spatial characteristics of an optical field. Unsurprisingly, they are used ubiquitously in the domain of structured light \cite{roadmap}. Of particular interest are the orbital angular momentum (OAM) spatial modes which are often touted as having the potential to enhance many key optical technologies \cite{OAM1}. In many applications, there is a prospective advantage to be gained for using OAM due to the large (countably infinite) state space on offer \cite{Yao2011}. In this regard, SLMs play a significant role in the realisation of this advantage due to the ease in which they can impart optical OAM onto light: one simply displays a hologram of the azimuthal phase $\exp(i\ell\phi)$ and this transformation ladders the OAM of the field's photons by a value of $\ell\hbar$. More generally, SLMs can be used to arbitrarily shape the OAM states of light \cite{Pinnell2019}.

A natural question to ask at this point, which has received little attention in the literature, is: what are the SLM's limits for generating and detecting structured light fields? In this work, we aim to showcase what happens when one pushes SLMs to their limits. We do this in the context of generating and detecting vortex beams carrying OAM and use a phase-only, liquid crystal on silicon SLM (Holoeye Pluto) as an example.  We show how the pixelated nature of the screen degrades the vortex mode quality as the topological charge $\ell$ is increased. In so doing, we successfully generate very high order OAM-containing beams, the largest achieved to date using such devices. We believe our work will be of interest to the large community who utilise SLMs for OAM mode generation, manipulation and detection.

\section{Problem description and considerations}
\begin{figure}[t]
    \centering
    \includegraphics[width=\textwidth]{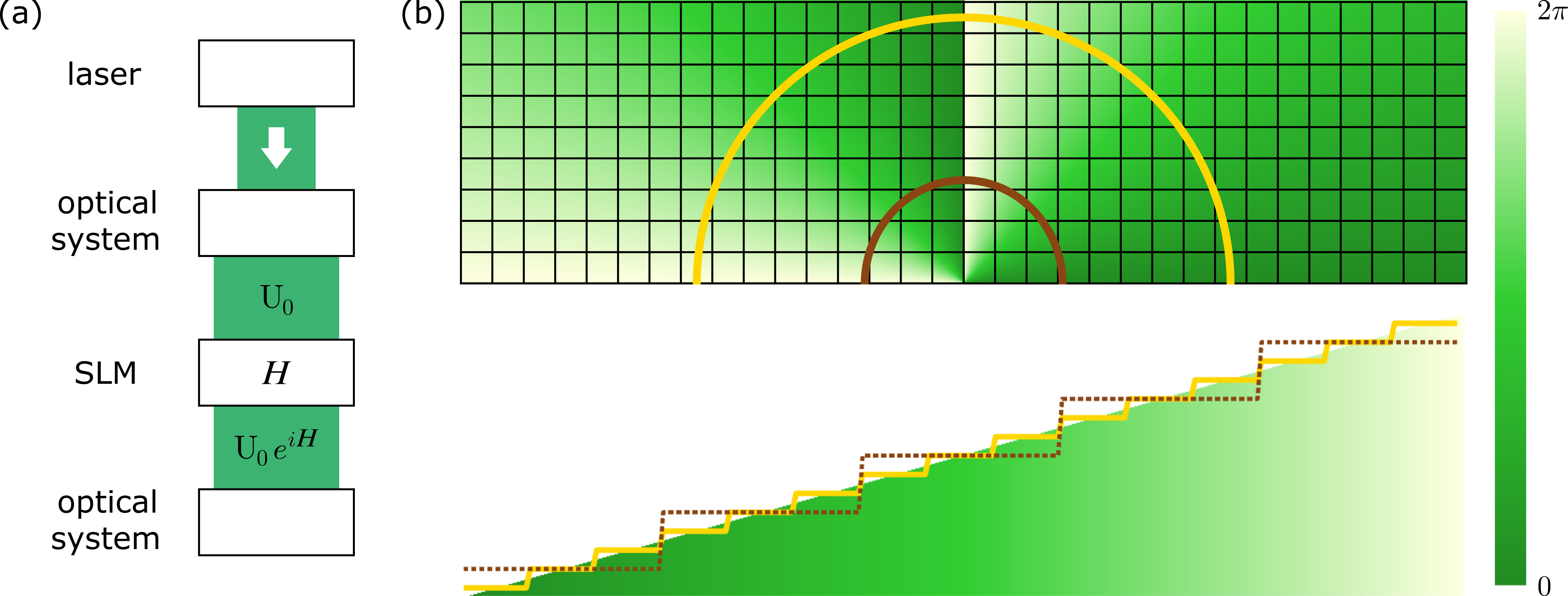}
    \caption{(a) Typical setup where SLMs are employed. Within an optical system, the SLM (in this case the phase-only variant) is used to digitally apply some transmission function $\exp(i H)$. (b) Due to the pixelated nature of the SLM screen, a phase function can only ever be approximated. For an azimuthal phase imparting OAM, this approximation worsens closer to the origin.}
    \label{fig:concept}
\end{figure}

We start by describing the general optical system we wish to consider, as depicted in Fig.~\ref{fig:concept}(a). Light emanating from a laser is modified and then guided towards an SLM by means of some optical system, typically a magnifying telescope. The optical field U$_0$ impinges onto the SLM which imparts a phase function that is determined by the hologram $H$ that is displayed on it, resulting in the modulated field U$_0\,e^{iH}$. More generally, SLMs impart some transmission function $T$ onto the optical field which is directly related to the displayed hologram $H$. Thereafter, the modified field is manipulated by another optical system which may include some form of detection and/or characterisation. 

In this work, we are mainly interested in studying OAM-containing fields generated by means of SLMs. For the particular case of phase-only SLMs, in order to impart an OAM of $\ell\hbar$ per photon onto the beam, one simply displays the phase function 
\begin{equation}
    H(\phi) = \ell \phi \mod 2\pi\,,
\end{equation}
where $\phi$ is the azimuthal polar coordinate. Since the SLM is not perfectly efficient, a diffraction grating must be added to separate the modulated and unmodulated light. Such a hologram can be written as,
\begin{equation} \label{eq:H}
    H(x_i,y_j) = \ell \, \tan^{-1} \frac{y_j}{x_i} + 2\pi\, G_x \, x_i + 2\pi\, G_y \, y_j \mod 2\pi\,,
\end{equation}
where $(x_i,y_j)$ are the (column, row) coordinates of the $(i,j)$th pixel on the screen and $(G_x,G_y)$ is the spatial frequency vector of the diffraction grating. Since Laguerre-Gaussian (LG) modes are the most natural basis for OAM-containing optical fields, we will modulate the hologram in Eq.~\ref{eq:H} with the annular LG amplitude function, given by,
\begin{equation} \label{eq:LGamp}
    \text{LG}_0^\ell(r) = \sqrt{\frac{2}{\pi |\ell|!}} \, \frac{1}{w_0}\, \left(\frac{\sqrt{2}r}{w_0}\right)^{|\ell|}\exp\left(-\frac{r^2}{w_0^2}\right)\,,
\end{equation}
where $w_0$ is the Gaussian waist radius. The standard procedure for generating an LG mode is to sufficiently expand the incoming laser beam so that it is approximately constant over the SLM screen \cite{SPIEbook}. Then, the beam shaping problem is reduced to finding the hologram that enacts the transmission function $T = \text{LG}_0^\ell(r)\exp(i\ell\phi)$. Several techniques exist for translating such a function into a phase-only hologram, but since high fidelity modes are desired we will utilise an exact encoding procedure as detailed in Ref.~\cite{Bolduc2013}. 

We then wish to quantify what happens to the resulting vortex field emanating from the SLM as the topological charge $\ell$ is increased. Is there a limiting value? What aspects of the implementation constrain such a value? It turns out that there are several considerations that need to be addressed in order to answer these questions, which can be categorised into physical, technological and numerical. In what follows, we will elaborate on these in turn.

\subsection{Physical considerations}
\noindent For the SLM to be effective in detecting and/or manipulating an OAM mode, it is an obvious requirement that the entire beam should fit onto the SLM screen. We will now compute an approximate theoretical limit for the highest order vortex beam that can fit onto the SLM. A good starting point is to calculate the maximum OAM that a propagating vortex beam with an azimuthal phase $\exp (i\ell\phi)$ can have within a given optical system. It is straightforward to verify that there is an inevitable optical vortex density limit within a circular region of radius $R$ (centred on the optical axis) which is given by \cite{roux2003oamlimit},
\begin{equation} \label{eq:oamLimit}
     |\ell|_{\text{max}} = \frac{2 \pi R \,\text{NA}}{\lambda} \,,
\end{equation}
where NA is the numerical aperture of the optical system and $\lambda$ is the light's wavelength. For vortex modes with $|\ell| > |\ell|_\text{max}$, evanescent waves are excited within the circular region and so the transverse amplitude there decays to zero over a length on the scale of the wavelength. If $R$ corresponds to the minimum aperture of the optics, these vortex modes will not propagate through the system.

In the case of requiring an OAM mode to fit onto the SLM so that it can be detected and/or manipulated, it is natural to say that the evanescent region (the vortex core) of the OAM mode should not be larger than the SLM. If the screen has pixel dimensions $N_x \times N_y$, each with a pitch of $(\Delta_x,\Delta_y)$, the largest circle that can be inscribed on it has radius,
\begin{equation}
    R = \frac{1}{2} \min (N_x \Delta x, N_y \Delta_y) \,.
\end{equation}
Note that the actual maximum vortex radius will be somewhat smaller than this since there should still be room on the screen for the OAM ``doughnut'' itself which is situated at a radius larger than the vortex core. Substitution of this radius and of the parameters of the given optical system into Eq.~\ref{eq:oamLimit} yields the absolute upper bound on the largest OAM mode that can physically be manipulated using an SLM.

Considering our SLM with pixel dimensions $1920 \times 1080$ each having a pitch of $8 \, \mu \text{m}$ and a He-Ne laser source operating at $\lambda = 633 \, \text{nm}$, the upper bound corresponds to,
\begin{equation}
    |\ell|_\text{max} \sim 40,000 \, \text{NA} \,.
\end{equation}
As discussed earlier, the displayed digital holograms are typically given a diffraction grating to separate the modulated and unmodulated light. To prevent the unmodulated light from propagating through the optical system, a pinhole/iris is used to block it. The inclusion of a pinhole can significantly limit the value of $\text{NA}$ and, by extension, $|\ell|_\text{max}$. In response, one can increase the frequency of the diffraction grating, thus widening the distance between the diffraction orders and allowing one to increase the size of the iris (increasing the numerical aperture). However, there is an inevitable trade-off since a grating frequency that is too large will degrade mode quality, due to the finite resolution and finite phase depth of the SLM screen. 

\subsection{Technological considerations}
\noindent If one could attempt to summarise the appeal of SLMs as generally as possible, it would be reasonable to say that the main attraction lies in its ability to digitally apply an arbitrary transmission function $T(\mathbf{x}) = A(\mathbf{x}) \exp[ i \Phi(\mathbf{x})]$ to an optical field. Since SLMs employing liquid crystal displays are phase-only devices, one has to determine an appropriate phase function (hologram) $H(\mathbf{x})$ that will execute the desired transmission function. Note that the equivalence of $\exp[i H(\mathbf{x})] = T(\mathbf{x})$ only holds in the continuum limit, whereas the actual displayed hologram on the SLM is discrete due to the small (but non-zero) pixel size. This is conceptually shown in Fig.~\ref{fig:concept}(b) for a hologram of the form $H(\phi) = \ell\phi$. What is the effect that this pixelation has on the output OAM mode? 

\begin{figure}[t]
    \centering
    \includegraphics[width=\textwidth]{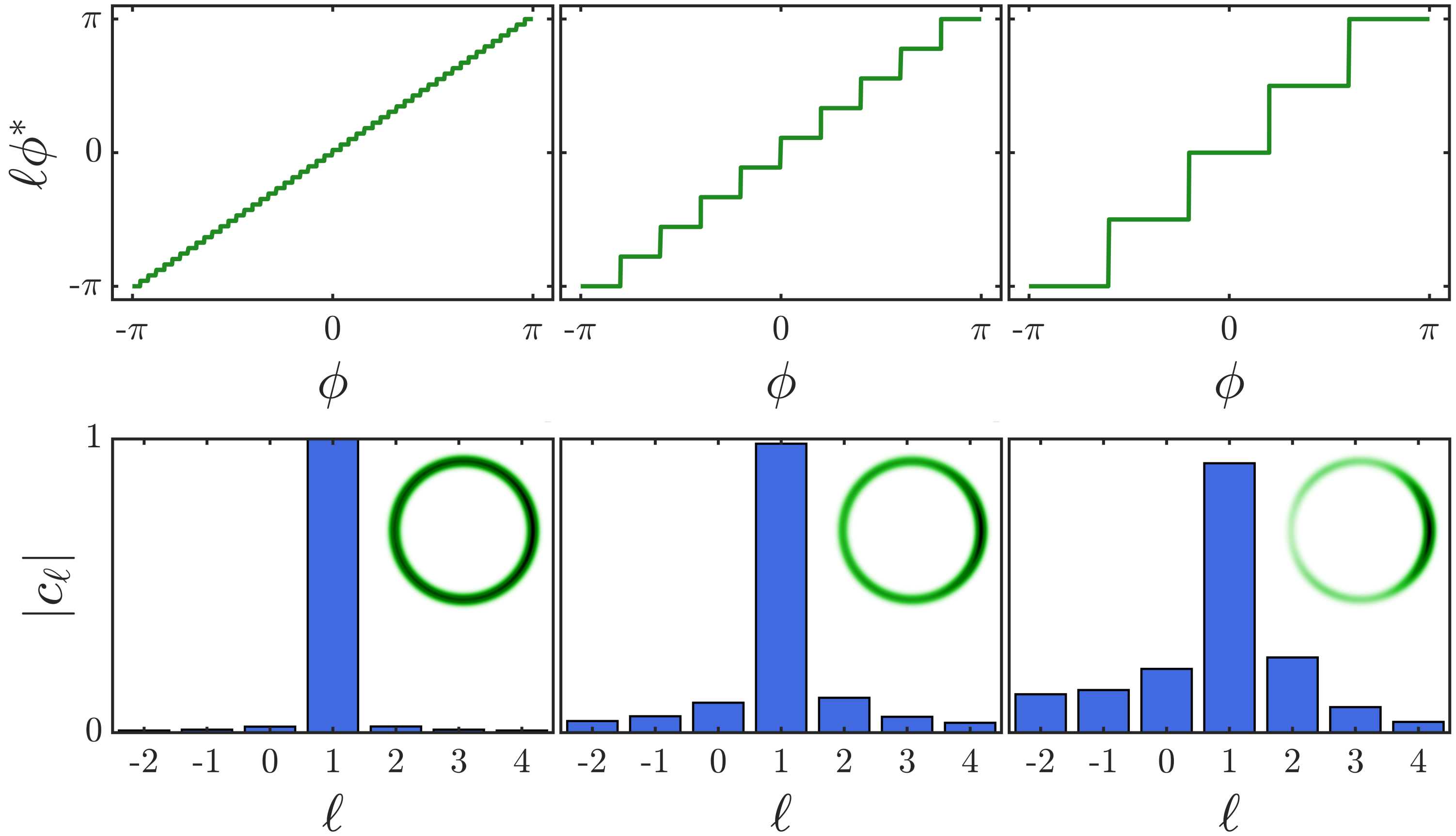}
    \caption{Simulation showing the decomposition of a discretised azimuthal phase ($\ell=1$) into the OAM basis. A more coarse discretisation results in more OAM mode cross-talk. Insets show the effect of the corresponding mode distribution on the beam when the carrier amplitude is an annulus. These effects should be similar to having a fixed SLM resolution and increasing the topological charge.}
 \label{fig:Simulation}
\end{figure}

To analytically investigate this in the context of vortex beams, we consider a step-function approximation of an azimuthal phase \cite{vijayakumar2019generation}
\begin{equation}
    \ell \phi^* = \sum_{k=0}^{N-1} \frac{2\pi k \ell}{N} \left[ \theta\left(\phi-\frac{2\pi k}{N}\right) - \theta\left(\phi - \frac{2\pi(k+1)}{N} \right) \right] \,,
\end{equation}
where $N$ is the number of steps and $\theta(\cdot)$ is the Heaviside step function. In Fig.~\ref{fig:Simulation}, we decomposed the approximate azimuthal phase $\exp(i\ell\phi^*)$  for $\ell=1$ in terms of the OAM eigenstates $\exp(i\ell\phi)$. It is evident that as the ``resolution'' of the approximation diminishes (as $N$ is decreased), more and more ``undesireable'' OAM modes are introduced into the field. This has a visible effect on the amplitude distribution of the beam, causing the intensity to be skewed to one side of the doughnut. 

Keeping the topological charge fixed and decreasing the resolution in this way should be similar to increasing the topological charge with a fixed resolution (as will be the scenario when we study this experimentally). This analysis leads us to hypothesise that attempting to generate vortex modes with larger $\ell$ values with SLMs would result in a larger spread of OAM modes and an increasingly non-uniform amplitude profile.

\subsection{Numerical considerations}
\noindent When probing the limits of structured light generation, an unexpected computational issue arises. Namely, the inability to compute the very large and very small numbers required for the generation of the digital hologram. We mention this fact here for completeness and expand on it in the context of generating LG modes with a large azimuthal (OAM) index $\ell$. Specifically, in
\begin{equation} \label{eq:LG}
    \text{LG}_0^\ell(r) \propto \underbrace{\left(\frac{\sqrt{2}r}{w_0}\right)^{|\ell|}}_{\rightarrow \infty}\overbrace{\exp\left(-\frac{r^2}{w_0^2}\right)}^{\rightarrow 0}\,,
\end{equation}
the radial terms clash: the $r^{|\ell|}$ term diverges to infinity whilst the Gaussian envelope (which ensures that the beam energy remains finite) necessarily becomes very small.
When performing complex amplitude modulation for the generation of the digital hologram, the above expression has to be evaluated and fed into an amplitude-modulating function (in our case a look-up table for the inverse of the sinc function). Even for modest values of $\ell$ (of the order of 100), these two radial terms approach the limit of the default floating point arithmetic in most software packages. In the following, we propose some workarounds.

The most obvious solution is to see if the user's software supports quadruple (or higher) precision. Many programming languages support or have packages for extending the precision of stored numbers. In our experience, we found Matlab's variable-precision functionality quite difficult to work with, especially for the necessarily large matrix sizes required for generating holograms. Mathematica was able to compute high-order LG amplitudes without having to load any specialised packages or call any specialised functions, but seems to be rather inefficient for large matrix sizes.


A simpler solution is to use a mode approximation, which necessarily becomes better in the regime of large mode indices. For example, a good approximation of Eq.~\ref{eq:LG} when $|\ell| \gg 1$ is,
\begin{equation} \label{eq:LGapprox}
    \text{LG}_0^\ell(r) \sim \exp \left( - \frac{\left( r - r_\ell \right)^2}{(w_0/\sqrt{2})^2} \right) \,.
\end{equation}
This formula makes use of the fact that the maximum amplitude of an LG mode occurs at the radius,
\begin{equation}
    r_\ell = \sqrt{\frac{|\ell|}{2}}w_0 \,.
\end{equation}
For $\ell \geq 100$, the error in the approximation is less than $0.01\%$. Note that when generating digital holograms, field amplitudes are eventually normalised to the interval $[0,1]$ so as to make full use of the available SLM phase depth. Hence, one does not need to be concerned with normalisation factors and the amplitude as specified by Eq.~\ref{eq:LGapprox} can be used. In any case, for large $\ell$, computing the normalisation factor for the true LG amplitude becomes intractable due to the $|\ell|!$ term.

Finally, the use of mathematical tricks can help to reduce the computational load. In the case of LG modes, taking the logarithm of the competing radial terms and exponentiating later can help to ensure that the computed values remain within the default numerical precision. Specifically, one computes the intermediate radial term,
\begin{equation}
    t_{rad} = |\ell| \log r - \frac{r^2}{w_0^2} \,.
\end{equation}
The LG amplitude is then precisely $\exp ( t_{rad} )$, up to a constant. This method is exact but will also eventually reach the software's default precision limit.

We note that the numerical procedure we consider for computing $H(\mathbf{x})$ given $T(\mathbf{x})$ requires the inversion of the sinc function, which can only be approximated numerically, having potentially undesirable consequences if the inversion contains significant errors. The numerical procedure for the inversion can also be computationally expensive, especially if this has to be done for all $\sim 10^6$ pixels. Thus, it is favourable to compute the inversion beforehand as accurately as is necessary and store the values in a look-up table, which can be applied to the screen array as a whole using vectorisation. Some research suggests that naive encoding (neglecting amplitude modulation) can produce similar results in some cases \cite{clark2016comparison}, but we recommend against this practice in general because it fundamentally does not encode the correct transmission function from the start. 

Finally, we note that any computational issues of the types described above fall away when the desired transmission function is phase-only since, in the case of vortex modes, one simply encodes $\exp(i \ell \phi)$ which can be done without computational difficulty.

\section{Experimental details}
\begin{figure}[t]
    \centering
    \includegraphics[width=\textwidth]{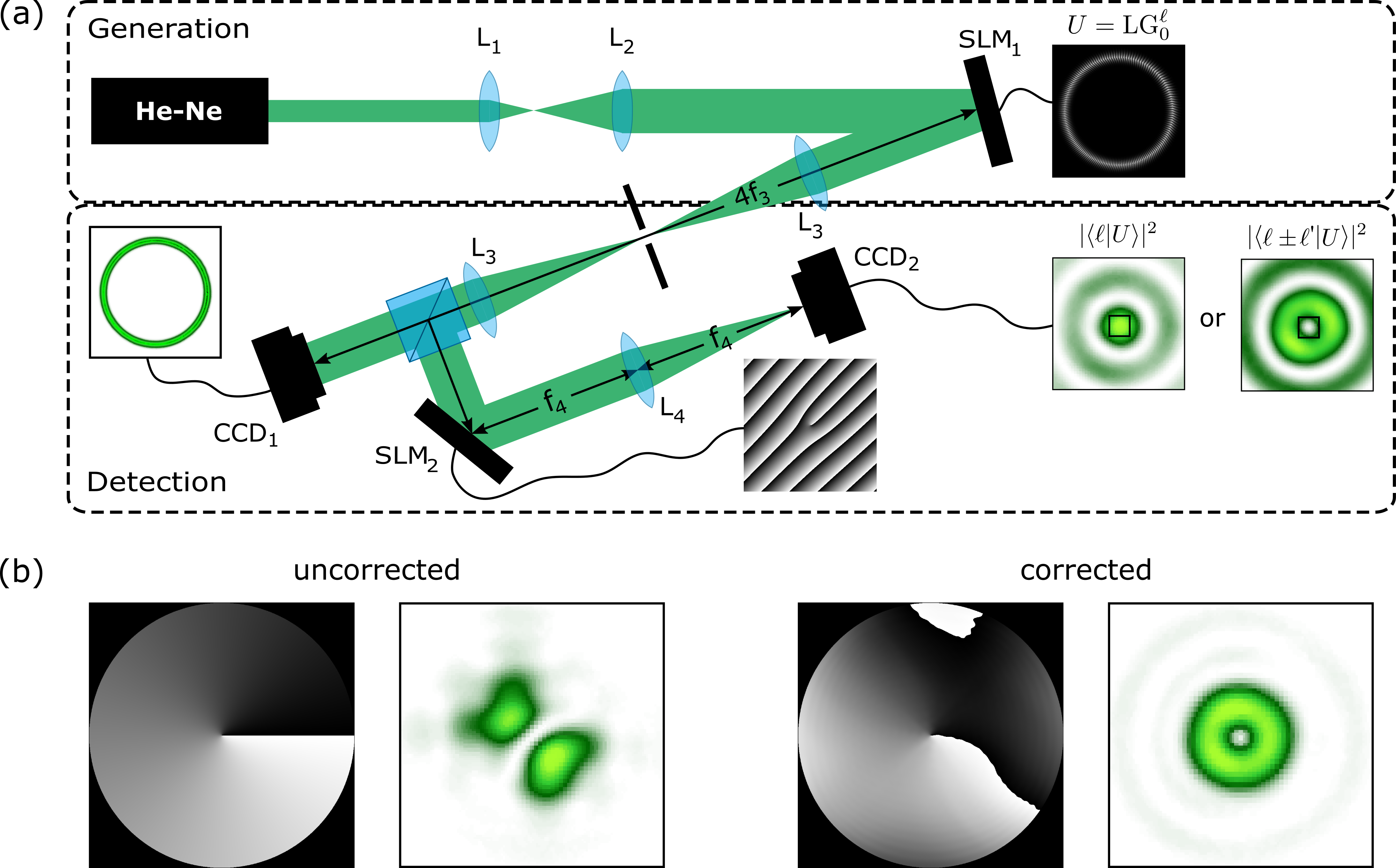}
    \caption{(a) Experimental setup used to test the limits of vortex mode generation and detection with SLMs. The first part of the setup generates the desired LG-OAM mode. The second part of the experiment simultaneously images the OAM mode to a CCD camera and performs a modal decomposition. (b) Wavefront correction of the SLM screen by applying the Gerchberg-Saxton algorithm to an image of a distorted $\ell=1$ vortex mode.}
    \label{fig:setup}
\end{figure}
\noindent Figure \ref{fig:setup}(a) depicts the optical setup built to probe the limits of vortex mode generation and detection using SLMs. The LG-OAM mode is carved out from an expanded and collimated He-Ne beam using the first SLM. This field is then imaged using a $4f$ lens system to the first CCD camera where amplitude information of the generated OAM mode is obtained. Simultaneously, using a beam splitter, the field is sent to the second SLM which, together with a lens and another CCD camera, performs a modal decomposition on the OAM content. From the resulting mode spectrum, we can glean information about the mode cross-talk induced by the SLM pixelation.

Although the SLM manufacturing process is technically sound, it is nevertheless imperfect which means that the screen is unlikely to be optically flat. The consequence is that mode quality will suffer, the effect of which will be more apparent as the mode propagates. It is reasonable to expect that the larger the portion of the screen the beam encounters, the larger the amount of induced wavefront aberrations. Therefore, an intuitive (and easily applied) heuristic to minimise wavefront errors after a beam encounters an SLM is to minimise the beam size. The lower bound is not zero, however, since as the beam size is made smaller the pixelated approximation of the transmission function encoded by the digital hologram is worsened (as discussed earlier). An experimenter could intuit an optimal beam size from experience based on their needs.

An easy way to circumvent such considerations is to apply wavefront correction to the SLM to make the screen optically flat. There exists a simple method for the correction of wavefront distortions of an SLM based on the application of the Gerchberg-Saxton algorithm to a single camera image of an optical vortex at the Fourier plane \cite{jesacher2007wavefront}. This method is favourable since it is mostly computational and does not require the use of an interferometer. A downside is the fact that this method can only correct for relatively small wavefront distortions, but this is not such a limitation since SLM wavefront distortions are likely to be small anyway. As can be seen in Fig.~\ref{fig:setup}(b), a significant improvement in vortex mode quality can be gained by applying wavefront correction. When performing modal decomposition, where one samples the beam intensity around the optical axis at the Fourier plane, this correction is almost compulsory to achieve optimal performance.


\section{Results}
\begin{figure}[t]
    \centering
    \includegraphics[width=\textwidth]{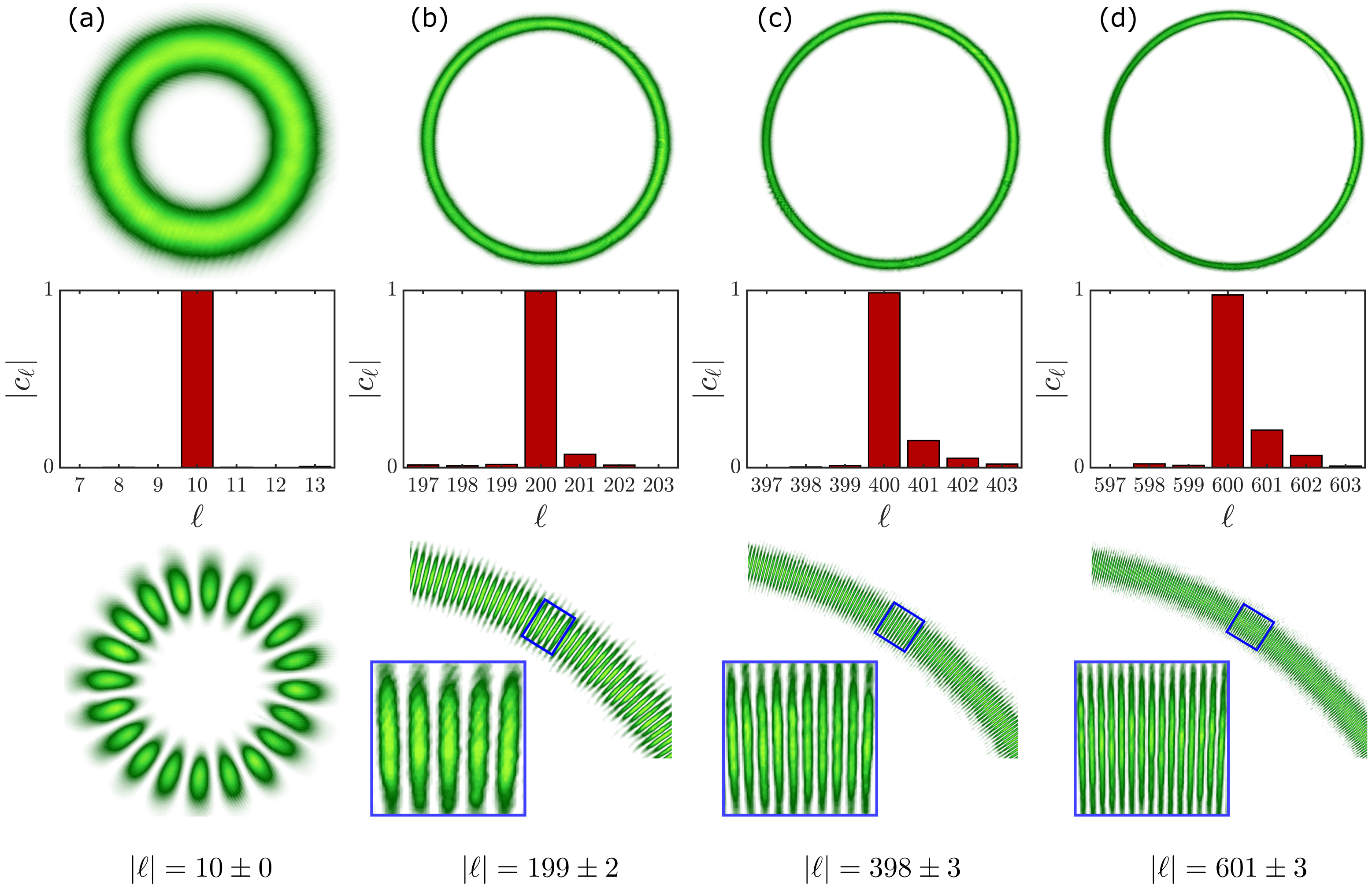}
    \caption{Camera images of generated OAM modes (first row) shown with the corresponding OAM mode spectrum (second row) for (a) $\ell = 10$, (b) $\ell = 200$, (c) $\ell=400$ and (d) $\ell = 600$ with $w_0 = w_0^L$. One can see the degradation of the OAM mode quality for larger topological charge values, as indicated by the increasingly non-uniform intensity and also the increasing mode cross-talk. Interferograms of the vortex mode interfered with its conjugate (third row) provide an independent qualitative measure of its OAM content.}
    \label{fig:ModeImages}
\end{figure}
\noindent On the first SLM, holograms corresponding to that depicted in Fig.~\ref{fig:concept}(b) with the LG amplitude envelope are displayed. The spatial frequency of the diffraction grating is kept constant and the topological charge of the LG-OAM mode is varied. In order to deduce the effect of pixelation on vortex mode quality, we would like for the beam size to remain fixed while its OAM content changes. Unfortunately, the doughnut radius of vortex modes increases with increasing $|\ell|$. This can be compensated for by scaling the embedded Gaussian waist radius ($w_0$ from Eq.~\ref{eq:LGamp}) as,
\begin{equation}
    w_0 \rightarrow \frac{w_0}{\sqrt{|\ell|+1}} \,.
\end{equation}
In essence, for any given value of $w_0$, this rescaling has the effect of fixing the second moment width of the LG annulus for all topological charges. For illustration, we consider three beam sizes: $w_0^S=1.4\, \text{mm}$ (small), $w_0^M=3.2\, \text{mm}$ (medium) and $w_0^L=5\,\text{mm}$ (large), the latter being the largest size possible for the particular SLM that we used. For each of the three sizes, the topological charge was steadily increased in increments of ten until the OAM density limit was reached (in accordance with Eq.~\ref{eq:oamLimit}). For each value of $\ell$, the intensity information was captured using CCD$_1$ and a modal decomposition \cite{Flamm2012B} was performed. An example of this is shown in the first two rows of Fig.~\ref{fig:ModeImages} for the largest beam size $w_0^L$. 

While many techniques exist for qualitatively deducing the topological charge of vortex beams, such as mode sorters \cite{Berkhout2010} and triangular apertures \cite{Melo2018}, these techniques are unable to provide full information of the modal content in the same way that modal decomposition can. In general, performing an OAM mode decomposition requires the application of a conjugate azimuthal phase (via some spiral-phase optic) and measuring the resulting on-axis intensity at the Fourier plane. In the absence of a practical and SLM-independent alternative for implementing modal decomposition, we decided to characterise the OAM modes generated by the SLM by using another SLM. Although not ideal, we favoured this approach to take advantage of the very well understood implementation of modal decomposition using SLMs \cite{Pinnell2020MDtut}. Nevertheless, in order to independently confirm the OAM content of the generated vortex beams, we built a simple interferometer after SLM$_1$ to interfere the generated vortex beam with its conjugate (achieved by having an odd number of mirrors in one arm and an even number in the other arm). The resulting interferogram is composed of $2|\ell|$ ``petals''. These petal modes were then magnified and segments of the beam were captured using a CCD camera. We wrote a simple image processing script to determine the arc angle of the captured doughnut segment and count the petals therein, whereupon the total number of petals (and hence $|\ell|$) can be estimated. This is shown in the third row of images in Fig.~\ref{fig:ModeImages}; we see that this qualitative measure supports the measured modal content using the SLM. We highlight that Fig.~\ref{fig:ModeImages}(d) corresponds to a vortex mode with a topological charge of $\ell = 600$ which is the largest value generated and detected with SLMs that has been reported to date.
\begin{figure}[t]
    \centering
    \includegraphics[width=0.8\textwidth]{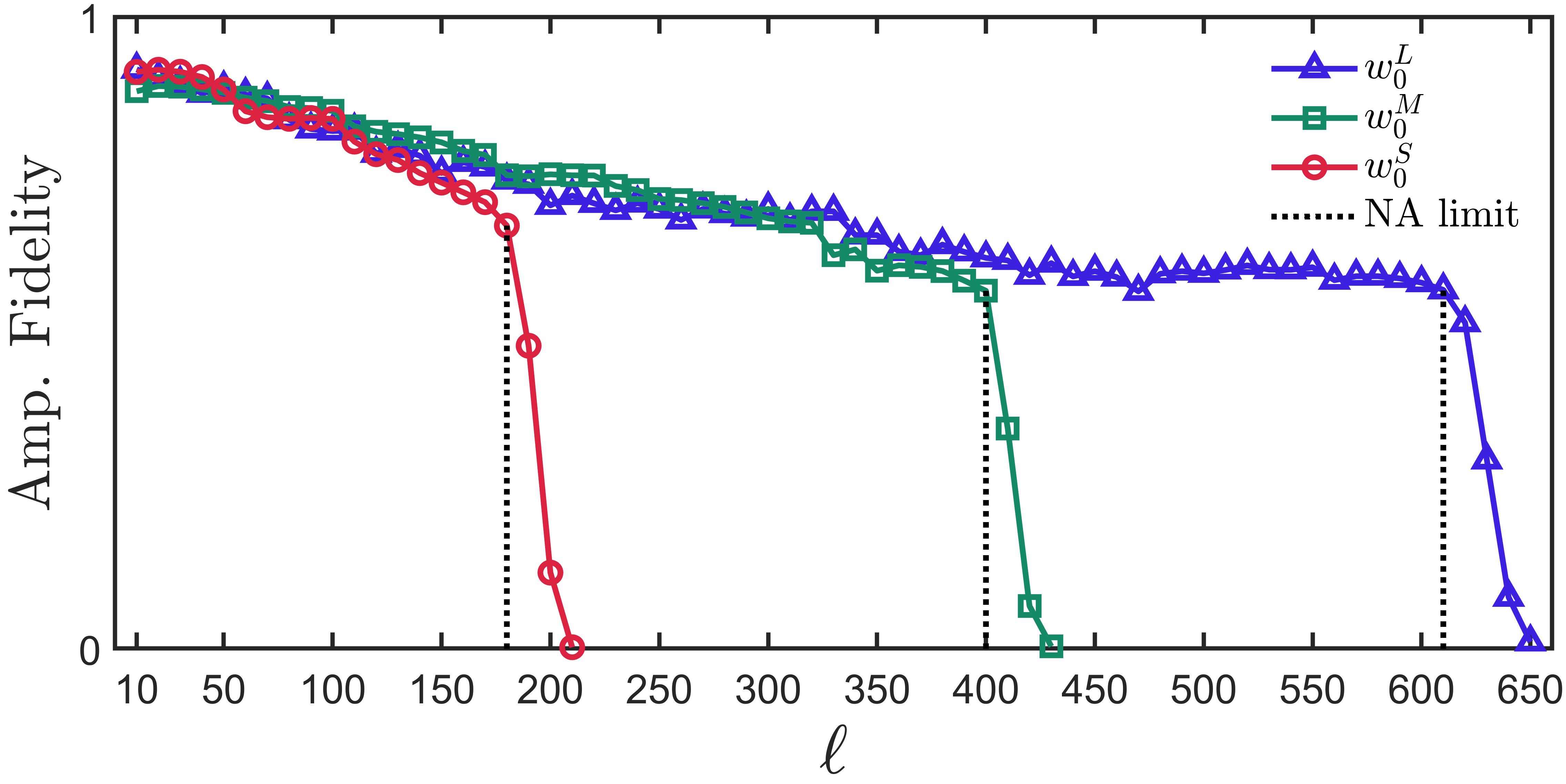}
    \caption{Amplitude fidelity of the generated vortex modes versus topological charge $\ell$ for the three beam sizes under consideration. Vertical dashed lines represent the attainment of the OAM density limit whereupon the mode fails to propagate through the optical system.}
    \label{fig:AmpFid}
\end{figure}

Figure \ref{fig:AmpFid} shows the results of the OAM amplitude fidelity as a function of topological charge for the three different beam sizes. This quantity was calculated using a strong image quality measure \cite{wang2002universalImage} that compares the observed beam intensity with the ideal intensity. The degradation of the amplitude for larger $\ell$ values is clearly visible; one can also see this effect visually in the top row of images in Fig.~\ref{fig:ModeImages}. Once the OAM density limit is reached, which is modest owing to the relatively small numerical aperture, more and more of the OAM mode falls within the evanescent region until the entire beam no longer propagates through the optical system. One would expect that OAM mode quality would degrade more rapidly for smaller beam sizes since there are fewer pixels available to effectively represent the azimuthal phase term of the transmission function $\text{LG}(r)\exp(i\ell\phi)$. The difference is not significant, but from Fig.~\ref{fig:AmpFid} it would appear that the linear decay in amplitude fidelity is slightly sharper for $w_0^S$ than it is for $w_0^L$.

\begin{figure}[t]
    \centering
    \includegraphics[width=0.8\textwidth]{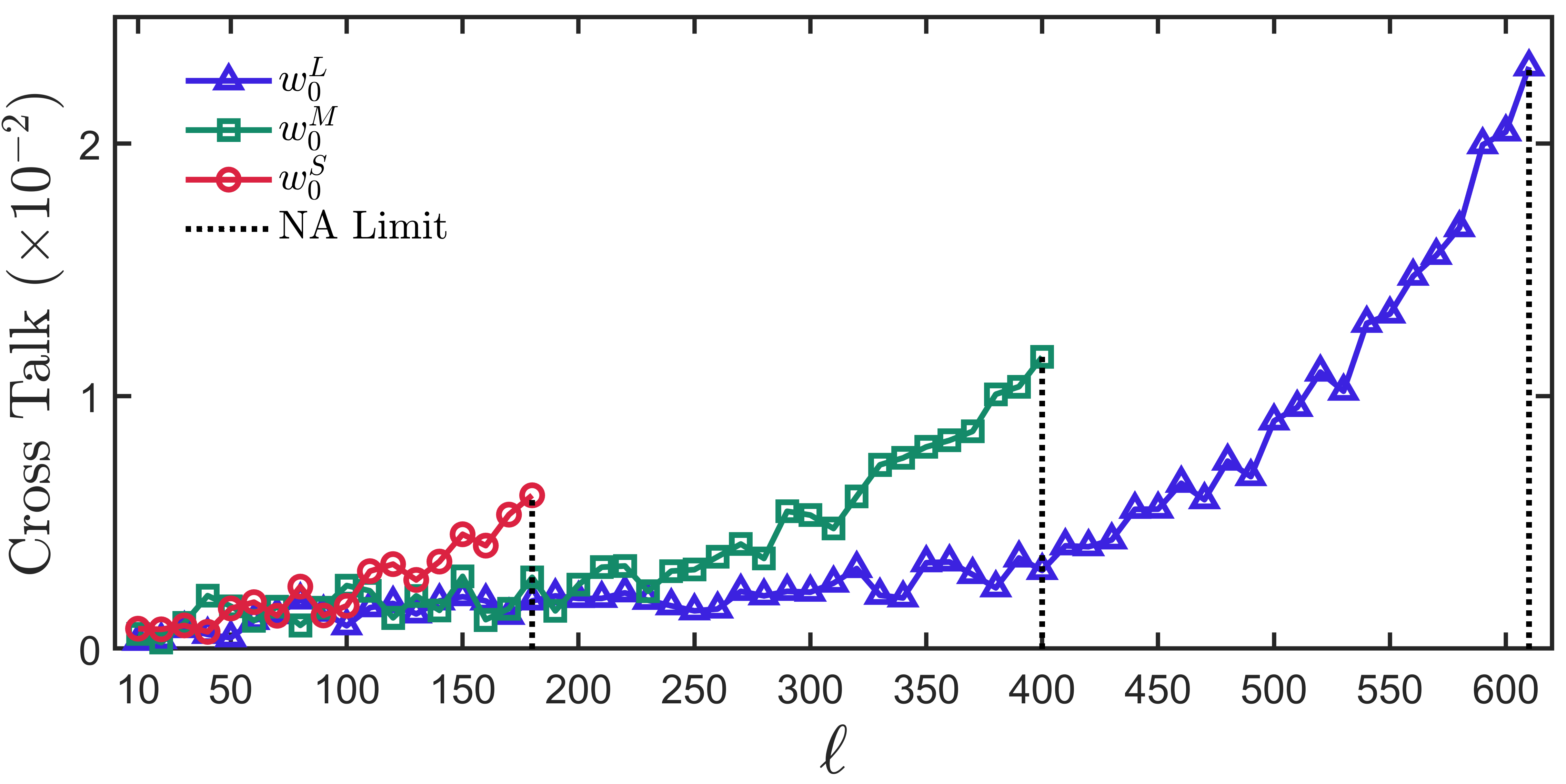}
    \caption{Mode cross-talk versus topological charge $\ell$ for the three beam sizes. The cross-talk is calculated from the sum of off-diagonal OAM components in the modal decomposition.}
    \label{fig:CrossTalk}
\end{figure}

A more quantitative measure of OAM mode quality is the cross-talk between modes. Figure \ref{fig:CrossTalk} shows the results of the OAM cross-talk as a function of topological charge for the three beam sizes. This quantity was calculated from the sum of the ``undesirable'' modal coefficients in the measured spectrum. The increasing cross-talk can also be observed in the second row of images of Fig.~\ref{fig:ModeImages}. It is clear from these results that OAM mode cross-talk increases in an exponential manner as $\ell$ is increased. Figure \ref{fig:CrossTalk} illustrates more clearly the fact that OAM mode quality degrades more rapidly for smaller beam sizes relative to the SLM size. Importantly, it also indicates that using as much of the SLM's active area as possible will maximise mode quality, provided that residual wavefront distortions are compensated for.

Although we do not explicitly consider them here, we anticipate that a similar analysis of vortex modes generated via Digital Micromirror Devices (DMDs) would exhibit similar characteristics.

\section{Conclusion}
\noindent In summary, we have considered the effect of screen pixelation on the quality of vortex modes generated using SLMs. Ultimately, and as expected, screen resolution primarily dictates the quality of modes that are accessible to experimenters who utilise SLMs for generation, manipulation and detection of structured light. In the context of generating vortex modes, we found that the optical system is perhaps the largest hindrance in the attainment of very high-order OAM modes. In spite of this, we successfully generated and detected OAM modes with a large topological charge of $\ell = 600$, which is significantly higher than what is previously reported. 

\section*{Acknowledgements}
J.P. acknowledges financial support from the Department of Science and Technology (South Africa).

\bibliography{mypaperdatabase}

\end{document}